\documentclass[twocolumn,aps]{revtex4}%
\usepackage{amsfonts}
\usepackage{amsmath}
\usepackage{amssymb}
\usepackage{graphicx}%
\begin{document}
\title{Negative Index Lens Aberrations}
\author{D. Schurig and D.R. Smith}
\affiliation{Physics Department, University of California, San Diego, La Jolla, CA, 92093}

\begin{abstract}
We examine the Seidel aberrations of thin spherical lenses composed of media
with refractive index not restricted to be positive. We find that
consideration of this expanded parameter space allows reduction or elimination
of more aberrations than is possible with only positive index media. In
particular we find that spherical lenses possessing real aplanatic focal
points are possible only with negative index. We perform ray tracing, using
custom code that relies only on Maxwell's equations and conservation of
energy, that confirms the results of the aberration calculations.

\end{abstract}
\date{\today}
\maketitle

\bigskip

In 1968, V. G. Veselago proposed the idea that a material could have a
negative index of refraction, and described how this would impact many basic
electromagnetic phenomena\cite{veselago}. In recent years, there has been
great interest in this subject due to experimental demonstrations of negative
index artificial materials\cite{smithPRL}, and the introduction of the perfect
lens concept\cite{perfectLens}. A perfect lens is a \emph{flat} slab of index
minus one, which can focus images with resolution exceeding that possible with
positive index optics. 

Recently, focusing by a negative index medium with \emph{curved} surfaces has
been experimentally demonstrated\cite{nimlens}. Traditional spherical profile
lenses composed of negative index media have several advantages over their
positive index counterparts: they are more compact, they can be perfectly
matched to free space, and here we demonstrate that they can also have
superior focusing performance.

The monochromatic imaging quality of a lens can be characterized by the five
Seidel aberrations: spherical, coma, astigmatism, field curvature and
distortion. These well known corrections to the simple Gaussian optical
formulas are calculated from a fourth order expansion of the deviation of a
wave front from spherical. (A spherical wave front converges to an ideal point
focus in ray optics). The coefficients in this expansion quantify the
non-ideal focusing properties of an optical element for a given object and
image position\cite{mahajan}. We find that there is an asymmetry of several of
the Seidel aberrations with respect to index about zero. Considering that an
interface with a relative index of +1 is inert and one of relative index -1 is
strongly refractive, this asymmetry is not surprising. However, our conclusion
that the asymmetry can yield superior focusing properties for negative index
lenses is not obvious.

We note that negative index media are necessarily frequency dispersive, which
implies increased chromatic aberration and reduced bandwidth. However,
diffractive optics, which possess a similar limitation, have found utility in
narrow band applications\cite{diffractiveOptics}.

To confirm the analytical aberration results, we developed custom ray tracing
code that does not rely on the sign of the index to determine the path of the
ray, but relies only on the permittivity, $\varepsilon$, the permeability,
$\mu$, Maxwell's equations and conservation of energy.

Between interfaces, in homogenous media, the ray propagates in a straight line
following the direction of the Poynting vector. Refraction across an
interface, from a region labeled $1$ into a region labeled $2$, is handled as
follows. Wave solutions are sought that satisfy the dispersion relation
(obtained from Maxwell's equations) in region $2,$%
\begin{equation}
\frac{c^{2}}{\omega^{2}}\mathbf{k}_{2}\cdot\mathbf{k}_{2}=\varepsilon_{2}%
\mu_{2},
\end{equation}
where $\mathbf{k}_{2}$ is the wave vector in region $2$. The solutions must
also satisfy a boundary match to the incident wave, requiring
\begin{equation}
\mathbf{n}\times\left(  \mathbf{k}_{2}-\mathbf{k}_{1}\right)  =0,
\end{equation}
where $\mathbf{n}$ is the unit normal to the interface. The outgoing,
refracted, wave must carry energy away from the surface if the incident wave
carried energy in,%
\begin{equation}
\left(  \mathbf{P}_{2}\cdot\mathbf{n}\right)  \left(  \mathbf{P}_{1}%
\cdot\mathbf{n}\right)  >0,
\end{equation}
where $\mathbf{P}=\frac{1}{2}\operatorname{Re}\left(  \mathbf{E}%
\times\mathbf{H}^{\ast}\right)  $ is the time averaged Poynting vector.
Finally, the wave must not be exponentially growing or decaying,
$\operatorname{Im}\left(  \mathbf{k}_{2}\right)  =0$, since the media are
assumed passive and lossless. If a solution exists that satisfies all the
above criteria, the ray is continued with the new found wave vector and
Poynting vector. Furthermore, since we consider only isotropic media the
solution will be unique.

We find that the form of the expressions for the Seidel aberrations of thin
spherical lenses found in the optics literature are unchanged by the
consideration of negative index media. We reached this conclusion by
re-deriving these expressions, from first principles, using only the
definition of optical path length and Fermat's Principle. We interpret the
optical path length, $OPL=\int_{C}n(s)ds$, to be the phase change (in units of
free space wavelength) that a wave would undergo along the path $C$, if $C$ is
oriented parallel to the Poynting vector. The optical path may have
contributions that are negative where the Poynting vector and the wave vector
are antiparallel, i.e. where the index is negative.\ These aberration formulae
are further corroborated by agreement with the results of our ray tracing.%
%TCIMACRO{\FRAME{ftbFU}{2.6238in}{0.8951in}{0pt}{\Qcb{Construction used for
%aberration calculation. The aperture stop, labeled AS is at the plane of the
%thin lens (though lens shown is thick). The Gaussian image plane is labeled
%IP. The aperture stop coordinate vector, $\mathbf{r}$, and the image plane
%coordinate vector, $\mathbf{h}$, are not necessarily parallel as shown.}%
%}{\Qlb{fig aberration diagram}}{fig0.eps}%
%{\special{ language "Scientific Word";  type "GRAPHIC";
%maintain-aspect-ratio TRUE;  display "ICON";  valid_file "F";
%width 2.6238in;  height 0.8951in;  depth 0pt;  original-width 3.2932in;
%original-height 1.0629in;  cropleft "0";  croptop "1";  cropright "1";
%cropbottom "0";  filename 'fig0.eps';file-properties "XNPEU";}}}%
%BeginExpansion
\begin{figure}
[tb]
\begin{center}
\includegraphics[
height=0.8951in,
width=2.6238in
]%
{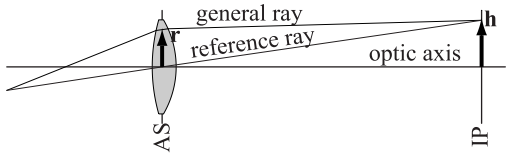}%
\caption{Construction used for aberration calculation. The aperture stop,
labeled AS is at the plane of the thin lens (though lens shown is thick). The
Gaussian image plane is labeled IP. The aperture stop coordinate vector,
$\mathbf{r}$, and the image plane coordinate vector, $\mathbf{h}$, are not
necessarily parallel as shown.}%
\label{fig aberration diagram}%
\end{center}
\end{figure}
%EndExpansion

The wave aberration, $\Delta OPL$, is the difference in optical path length of
a general ray and a reference ray, where the reference ray passes through the
optic axis in the aperture stop and the general ray is parameterized by its
coordinate in the aperture stop, $\mathbf{r}$, and its coordinate in the image
plane, $\mathbf{h}$ (Fig. \ref{fig aberration diagram}). To be in the Gaussian
optic limit, where spherical interfaces yield perfect imaging, $r$ and $h$
must be near zero. A series expansion of the wave aberration in these
parameters
\begin{equation}
\Delta OPL=\sum_{l,m,n=0}^{\infty}C_{lmn}\left(  \mathbf{r}\cdot
\mathbf{r}\right)  ^{l}\left(  \mathbf{r}\cdot\mathbf{h}\right)  ^{m}\left(
\mathbf{h}\cdot\mathbf{h}\right)  ^{n}%
\end{equation}
yields corrections to Gaussian optics of any desired order. The lowest order
corrections for a thin spherical lens with aperture stop in the plane of the
lens are given by%

\begin{subequations}
\label{eq aberration coeffs}%
\begin{align}
C_{200}  &  =-\frac{1}{32f^{\prime3}n\left(  n-1\right)  ^{2}}\times
\nonumber\\
&  \left[  n^{3}+\left(  n-1\right)  ^{2}\left(  3n+2\right)  p^{2}+4\left(
n+1\right)  pq+\left(  n+2\right)  q^{2}\right]  ,\\
C_{110}  &  =-\frac{1-p}{8f^{\prime3}n\left(  n-1\right)  }\left[  \left(
2n+1\right)  \left(  n-1\right)  p+\left(  n+1\right)  q\right]  ,\\
C_{020}  &  =-\frac{\left(  1-p\right)  ^{2}}{8f^{\prime3}},\\
C_{101}  &  =-\frac{\left(  1-p\right)  ^{2}}{16f^{\prime3}n}\left(
n+1\right)  ,\\
C_{011}  &  =0.
\end{align}
These coefficients are the Seidel aberrations: spherical, coma, astigmatism,
field curvature and distortion respectively. Also appearing in these
expressions are $p$, the position factor, and $q$, the shape factor, where we
follow the definitions of Mahajan\cite{mahajan}. The position factor is given
by
\end{subequations}
\begin{equation}
p\equiv1-\frac{2f^{\prime}}{S^{\prime}},
\end{equation}
where $f^{\prime}$ is the focal length referred to the image side and
$S^{\prime}$ is the image position. Through the thin spherical lens imaging
equation,%
\begin{equation}
\frac{1}{S^{\prime}}-\frac{1}{S}=\frac{1}{f^{\prime}}=\left(  n-1\right)
\left(  \frac{1}{R_{1}}-\frac{1}{R_{2}}\right)  ,
\end{equation}
where $S$ is the object position and $R_{1}$ and $R_{2}$ are the lens radii of
curvature, the position factor is directly related to the magnification,%
\begin{equation}
M=\frac{S^{\prime}}{S}=\frac{p+1}{p-1}.
\end{equation}
The shape factor is given by%
\begin{equation}
q\equiv\frac{R_{2}+R_{1}}{R_{2}-R_{1}}%
\end{equation}
A lens with a shape factor of 0 is symmetric, and $\pm1$ is a plano-curved
lens. Using the shape and position factor, all thin spherical lens
configurations are described.

We will first examine the very important case of a source object at infinite
distance. This is a position factor of $-1$. We are left with two parameters
that can be used to reduce aberrations, $n$ and $q$. We will set the value of
$q$ to eliminate one of the aberrations and compare the remaining aberrations
as a function of index. We will restrict our attention to moderate values of
index. At large absolute values of index, the aberrations approach the same
value independent of sign, but dielectric lenses with high index have
significant reflection coefficients due to the impedance mismatch to free
space.%
%TCIMACRO{\FRAME{ftbFU}{195.5625pt}{340.0625pt}{0pt}{\Qcb{Top plot shows
%spherical aberration (black), astigmatism (green), field curvature (blue), and
%shape factor (light gray) as a function of index for a lens focusing an object
%at infinity and bent for zero coma. Thin gray vertical lines indicate
%properties for lenses shown in ray tracing diagrams (bottom), meridional
%profile (left) and image spot (right). Incident angle is 0.2 radians and
%lenses are f/2. Index, shape factor, relative rms spot size, and spot diagram
%zoom are shown tabularly. In meridional profile, lens principle planes are
%shown as thin black vertical lines, and optic axis and Gaussian image plane
%are shown as blue lines. In spot diagram, Gaussian focus is at the center of
%blue cross hairs.}}{\Qlb{fig zero coma}}{fig1p4.eps}%
%{\special{ language "Scientific Word";  type "GRAPHIC";
%maintain-aspect-ratio TRUE;  display "ICON";  valid_file "F";
%width 195.5625pt;  height 340.0625pt;  depth 0pt;  original-width 3.2733in;
%original-height 7.0595in;  cropleft "0";  croptop "1";  cropright "1";
%cropbottom "0";  filename 'fig1p4.eps';file-properties "XNPEU";}}}%
%BeginExpansion
\begin{figure}
[tb]
\begin{center}
\includegraphics[
height=340.0625pt,
width=195.5625pt
]%
{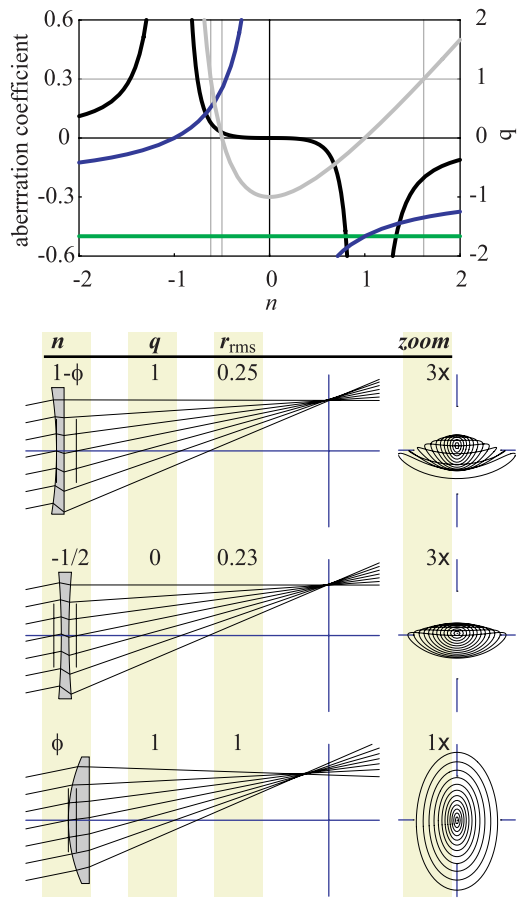}%
\caption{Top plot shows spherical aberration (black), astigmatism (green),
field curvature (blue), and shape factor (light gray) as a function of index
for a lens focusing an object at infinity and bent for zero coma. Thin gray
vertical lines indicate properties for lenses shown in ray tracing diagrams
(bottom), meridional profile (left) and image spot (right). Incident angle is
0.2 radians and lenses are f/2. Index, shape factor, relative rms spot size,
and spot diagram zoom are shown tabularly. In meridional profile, lens
principle planes are shown as thin black vertical lines, and optic axis and
Gaussian image plane are shown as blue lines. In spot diagram, Gaussian focus
is at the center of blue cross hairs.}%
\label{fig zero coma}%
\end{center}
\end{figure}
%EndExpansion%
%TCIMACRO{\FRAME{ftbFU}{198.6875pt}{350.125pt}{0pt}{\Qcb{All as in Fig.
%\ref{fig zero coma}, except the following. Lens is bent for zero spherial
%aberration. Coma is shown red. Solid and dashed lines indicate different
%solutions. Spot size, $r_{rms}$, is relative to bottom lens spot in Fig.
%\ref{fig zero coma}. All spot diagrams are at the same scale. }}{\Qlb{fig zero
%spherical}}{fig2p4.eps}{\special{ language "Scientific Word";
%type "GRAPHIC";  maintain-aspect-ratio TRUE;  display "ICON";
%valid_file "F";  width 198.6875pt;  height 350.125pt;  depth 0pt;
%original-width 3.2863in;  original-height 7.2973in;  cropleft "0";
%croptop "1";  cropright "1";  cropbottom "0";
%filename 'fig2p4.eps';file-properties "XNPEU";}}}%
%BeginExpansion
\begin{figure}
[tbtb]
\begin{center}
\includegraphics[
height=350.125pt,
width=198.6875pt
]%
{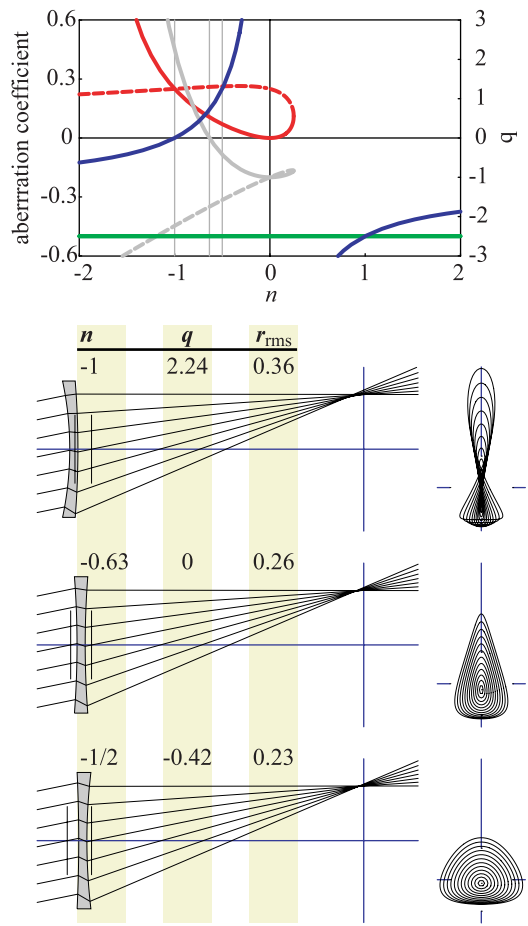}%
\caption{All as in Fig. \ref{fig zero coma}, except the following. Lens is
bent for zero spherial aberration. Coma is shown red. Solid and dashed lines
indicate different solutions. Spot size, $r_{rms}$, is relative to bottom lens
spot in Fig. \ref{fig zero coma}. All spot diagrams are at the same scale. }%
\label{fig zero spherical}%
\end{center}
\end{figure}
%EndExpansion%
%TCIMACRO{\FRAME{ftbFU}{196.6875pt}{402.5625pt}{0pt}{\Qcb{All as in Fig.
%\ref{fig zero coma}, except the following. Lens configuration with object and
%image at finite positions and bent for zero spherial aberration and coma.
%Position factor is shown dark gray. Real image object pairs only occur when
%position factor is in shaded region, $\left\vert p\right\vert <1$. Lens pairs
%are f/1.23, f/1.08, f/0.90 and have magnifications -1,-2,-3. In second to last
%spot diagram, horizontal (10\textsf{x}) and vertical (100\textsf{x}) zoom are
%not equal.}}{\Qlb{fig finite}}{fig3.eps}%
%{\special{ language "Scientific Word";  type "GRAPHIC";
%maintain-aspect-ratio TRUE;  display "ICON";  valid_file "F";
%width 196.6875pt;  height 402.5625pt;  depth 0pt;  original-width 3.314in;
%original-height 6.9323in;  cropleft "0";  croptop "1";  cropright "1";
%cropbottom "0";  filename 'fig3.eps';file-properties "XNPEU";}}}%
%BeginExpansion
\begin{figure}
[tbtbtb]
\begin{center}
\includegraphics[
height=402.5625pt,
width=196.6875pt
]%
{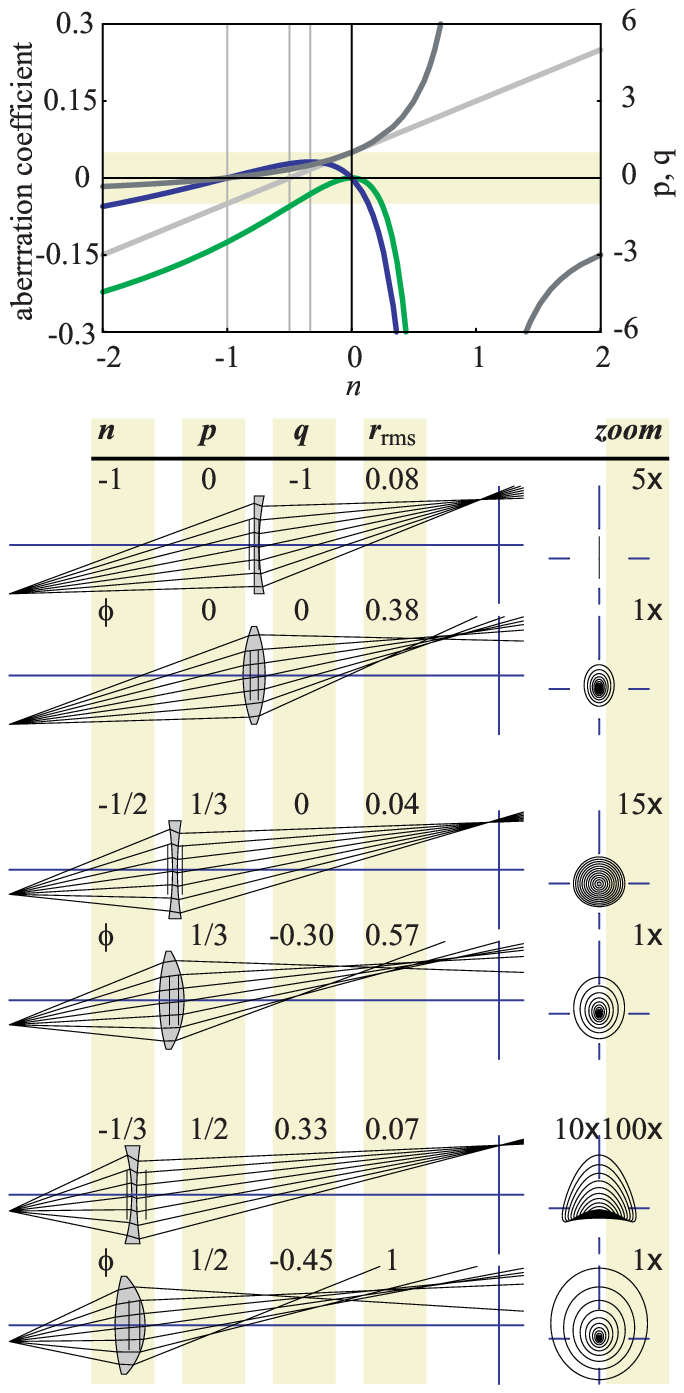}%
\caption{All as in Fig. \ref{fig zero coma}, except the following. Lens
configuration with object and image at finite positions and bent for zero
spherial aberration and coma. Position factor is shown dark gray. Real image
object pairs only occur when position factor is in shaded region, $\left\vert
p\right\vert <1$. Lens pairs are f/1.23, f/1.08, f/0.90 and have
magnifications -1,-2,-3. In second to last spot diagram, horizontal
(10\textsf{x}) and vertical (100\textsf{x}) zoom are not equal.}%
\label{fig finite}%
\end{center}
\end{figure}
%EndExpansion

The usual ordering of the aberrations is from highest to lowest in the order
of $r$, the aperture coordinate. This is the ordering of most image
degradation to least if one is forming images with significant lens aperture,
but small to moderate image size, which is a common occurrence in
applications. Thus, spherical aberration is an obvious target for elimination.
However, there are no roots of $C_{200}$ for values of index greater than one,
which is why this aberration is referred to as spherical aberration, since it
appears to be inherent to spherical lenses. The usual practice is to eliminate
coma (the next in line), and it so happens that the resulting lens has a value
for the spherical aberration that is very near the minimum obtainable.
Adjusting the shape factor, $q$, is often called lens bending. If we bend the
lens for zero coma, that is find the roots of $C_{110}$ with respect to $q$ we obtain%

\begin{equation}
q_{c}=\frac{\left(  2n+1\right)  \left(  n-1\right)  }{n+1}. \label{eq qc}%
\end{equation}
We plug this value for $q$ and $p=-1$ into (\ref{eq aberration coeffs}) and
plot the remaining three non-zero aberration coefficients as well as $q_{c}$
in Fig. \ref{fig zero coma}. We note that there are two values of index where
$q=1$, which represent a plano-concave/convex lens. Setting (\ref{eq qc}%
)\ equal to one we obtain,
\begin{equation}
n^{2}-n-1=0. \label{eq golden}%
\end{equation}
the roots of which are the ubiquitous golden ratios, $n=\phi\simeq1.62$ and
$n=1-\phi\simeq-0.62$\cite{goldenRatio}. We also note that there is a window
of index values near $n=-0.7$ where both the spherical aberration and field
curvature are small. There is no equivalent window in positive index.

Several ray tracing diagrams with both meridional rays and ray spot diagrams
are shown for specific values of index in Fig. \ref{fig zero coma}. The
reference lens has index $\phi$, which is close to typical values used in
visible optical lenses and near enough to $n=1$ for reasonably low reflection.
The lenses of negative index shown are in fact closer to $n=-1$, which is the
other index which permits perfect transmission, so this is a fair comparison.
The negative index lenses all show significantly tighter foci than the
positive index lens.

If we attempt to bend a lens with $p=-1$ to obtain zero spherical aberration
we obtain the two solutions%
\begin{equation}
q_{s}=\frac{2\left(  n^{2}-1\right)  \pm n\sqrt{1-4n}}{n+2}.
\end{equation}
These expressions have real values only for $n\leq1/4$, so an implementation
of such a lens (embedded in free space) is not possible with normal materials.
It is a surprising and significant result that negative index permits an
entire family of spherical aberration free spherical lenses that can focus a
distant object to a real focus, Fig. \ref{fig zero spherical}. The solution
with the negative sign in the expression for $q_{s}$ (solid curves) has less
coma for moderate negative values of index, so ray tracing diagrams are shown
for that solution. We note that at $n=-1$, the field curvature is also zero,
thus this lens has only two of the five Seidel aberrations, coma and
astigmatism. For a positive index reference we use the zero coma, $n=\phi$
lens from above. Here again, negative index lenses achieve a tighter focus
than a comparable positive index lens.

Now we examine the case of $|p|<1$, that is a real object and real image both
at finite position. Since $p$ and $q$ are both free parameters, we can
conceivably eliminate two aberrations. If we eliminate spherical aberration
and coma the resulting lens is called \emph{aplanatic}. It is a well known,
though incorrect, result that a spherical lens can can only have
\emph{virtual} aplanatic focal pairs. The correct statement is that only
negative index spherical lenses can have \emph{real} aplanatic focal pairs.

If we set $C_{200}$ and $C_{110}$ to zero and solve for $p$ and $q$, we obtain
four solutions, the two non-trivial ones are given by
\begin{subequations}
\label{eq pqsc}%
\begin{align}
p_{sc}  &  =\mp\frac{n+1}{n-1},\\
q_{sc}  &  =\pm\left(  2n+1\right)  .
\end{align}
We will focus on the solution with a minus sign for $p$ and the plus sign for
$q$. This solution has smaller aberrations for lens configurations that
magnify an image. The other solution is better for image reduction. Inserting
the expressions (\ref{eq pqsc})\ into (\ref{eq aberration coeffs}) we have
plotted the two remaining non-zero coefficient as well as the values of
$p_{sc}$ and $q_{sc}$ (Fig. \ref{fig finite}). Ray diagrams are shown for
lenses with magnifications of -1, -2 and -3. Also shown is a reference
positive index lens for each. The reference lenses (which cannot be aplanatic)
are of moderate index, $\phi$, with the same magnification and f/\# as the
lenses they are compared to. They are bent for zero coma but also have
spherical aberration near the minimum possible for the configuration. Again
the negative index lenses produce superior foci.

The lens of index $-1$ and magnification $-1$ is particularly interesting. At
this index value the field curvature is also zero. This remarkable lens
configuration has only one of the five Seidel aberrations, astigmatism. This
is confirmed by ray tracing which shows a one dimensional "spot" at the image
plane. This is perfect focusing in the sagittal plane. Perfect focusing also
occurs in the meridional plane, in front of sagittal focus.

One may ask why this asymmetric lens, $q=-1$, performs so well in a symmetric
configuration, $p=0$. This lens can be equivalently viewed as a biconcave
doublet with one component having index $-1$ and the other having index $1$,
i.e. free space. Driven by this observation, we found that all biconcave
doublets with arbitrary indices of $\pm n$ have identical focusing properties.
The only observable difference is in the internal rays, which are always
symmetric about the planer interface, but make more extreme angles at higher
index magnitude.

Fabrication of any of these negative index lenses is quite feasible using
periodically structured artificial materials. Current artificial material
designs can operate at frequencies from megahertz through
terahertz\cite{terahertz}, where there are numerous communication and imaging
applications. For example, lens antennas could benefit both by a reduction in
aberrations, which translates directly into increased gain, and by a reduction
of mass, afforded by low density artificial materials. Furthermore, these
lenses are even easier to implement than a perfect lens, since they lack its
severe structure period per wavelength requirements and are more tolerant to
losses\cite{criticalRoute}. Negative index lenses at visible light frequencies
may also be possible, by using photonic crystals, which have shown potential
for negative refraction\cite{ozbayPhotonicNegativeRefraction,sridharFlatLens}.

Using the current optical system design paradigm, aberrations are minimized by
combining elements with coefficients of opposite sign\cite{hecht}. However,
more elements mean greater complexity and cost. Taking advantage of an
expanded parameter space that includes negative index can reduce the number of
required elements---possibly even to one.

\bibliographystyle{apsrev}
\bibliography{acompat,dav}

\end{subequations}
\end{document}